\documentclass[journal]{IEEEtran}

\usepackage{anyfontsize} 
\usepackage{amsmath,amssymb,amsfonts}
\usepackage{array}
\usepackage[caption=false,font=normalsize,labelfont=sf,textfont=sf]{subfig}
\usepackage{textcomp}
\usepackage{stfloats}
\usepackage{url}
\usepackage{verbatim}
\usepackage{graphicx}
\usepackage{cite}
\usepackage{xcolor}
\usepackage{textcomp}
\usepackage{algorithm}
\usepackage{hyperref}
\usepackage{url}
\usepackage{booktabs}
\usepackage{multirow}
\usepackage{flushend}
\usepackage[nohyperlinks,nolist]{acronym}
\usepackage[normalem]{ulem}
\usepackage[capitalise]{cleveref}
\usepackage{pifont} 
\usepackage{xspace}
\usepackage[para]{threeparttable}

\usepackage{algorithm} 
\usepackage{algpseudocodex} 

\newcommand{\positenv}[2]{Posit$\langle #1, #2 \rangle$}
\newcommand{\positbits}[1]{Posit$#1$}

\newcommand{\ra}[1]{\renewcommand{\arraystretch}{#1}}

\newcommand{\YES}{\ding{51}\xspace} 
\newcommand{\NO}{\ding{55}\xspace} 

\begin{document}
\bstctlcite{bstctl:nodash}

\title{Digit-Recurrence Posit Division}

\author{
Raul~Murillo,
Julio~Villalba-Moreno,~\IEEEmembership{Senior~Member,~IEEE},
Alberto~A.~Del~Barrio,~\IEEEmembership{Senior~Member,~IEEE},
Guillermo~Botella,~\IEEEmembership{Senior~Member,~IEEE}
\thanks{Raul Murillo, Alberto A. Del Barrio and Guillermo Botella are with the Department of Computer Architecture and Automation, Faculty of Computer Science, Complutense University of Madrid, 28040 Madrid, Spain (e-mail: ramuri01@ucm.es; abarriog@ucm.es; gbotella@ucm.es).}
\thanks{Julio Villalba-Moreno is with the Department of Computer Architecture, Universidad de Málaga, 29071 Málaga, Spain (e-mail: jvillalba@uma.es).}
}


\IEEEpubid{This work has been submitted to the IEEE for possible publication. Copyright may be transferred without notice, after which this version may no longer be accessible.}

\maketitle

\begin{abstract}
Posit arithmetic has emerged as a promising alternative to IEEE 754 floating-point representation, offering enhanced accuracy and dynamic range. However, division operations in posit systems remain challenging due to their inherent hardware complexity. In this work, we present posit division units based on the digit-recurrence algorithm, marking the first implementation of radix-4 digit-recurrence techniques within this context. Our approach incorporates hardware-centric optimizations including redundant arithmetic, on-the-fly quotient conversion, and operand scaling to streamline the division process while mitigating latency, area, and power overheads. Comprehensive synthesis evaluations across multiple posit configurations demonstrate significant performance improvements, including more than 80\% energy reduction with small area overhead compared to existing methods, and a substantial decrease in the number of iterations. These results underscore the potential of our adapted algorithm to enhance the efficiency of posit-based arithmetic units.
\end{abstract}


\begin{acronym}[]
    \acro{AI}{artificial intelligence}
    \acro{ALU}{arithmetic logic unit}
    \acro{ASIC}{application-specific integrated circuit}
    \acro{BiCG}{biconjugate gradient}
    \acro{CISC}{complex instruction set computer}
    \acro{CNN}{convolutional neural network}
    \acro{CG}{conjugate gradient}
    \acro{DNN}{deep neural network}
    \acro{FPU}{floating-point unit}
    \acro{FMA}{fused multiply-add}
    \acro{FPGA}{field-programmable gate array}
    \acro{FF}{flip-flop}
    \acro{GPU}{graphics processing unit}
    \acro{GAN}{generative adversarial networks}
    \acro{GEMM}{general matrix multiplication}
    \acro{HDL}{hardware description language}
    \acro{HPC}{high-performance computing}
    \acro{HLS}{high-level synthesis}
    \acro{ISA}{instruction set architecture}
    \acro{IP}{intellectual property}
    \acro{LUT}{lookup table}
    \acro{LSB}{least significant bit}
    \acro{ML}{machine learning}
    \acro{MAC}{multiply-accumulate}
    \acro{MSE}{mean squared error}
    \acro{MSB}{most significant bit}
    \acro{MaxAbsE}{maximum absolute error}
    \acro{NaN}{Not a Number}
    \acro{NaR}{Not a Real}
    \acro{NN}{neural network}
    \acro{OS}{operating system}
    \acro{PAU}{posit arithmetic unit}
    \acro{RISC}{reduced instruction set computer}
    \acro{RTL}{register-transfer level}
    \acro{RMSE}{root mean squared error}
    \acro{SIMD}{single instruction, multiple data}
    \acro{ulp}{unit in the last place}

\end{acronym}


\section{Introduction}
\label{sec:introduction}

\IEEEPARstart{P}{osit} arithmetic~\cite{Gustafson2017Beating}, introduced in 2017, offers an alternative to the widely used IEEE 754 floating-point standard for representing and computing with real numbers. Posits deliver consistent results across different platforms and have minimal special cases. Additionally, posits inherently prevent overflow and underflow, simplifying exception handling.

Since the introduction of posits, the state of the art has become populated with several functional unit designs, mainly adders, multipliers and \ac{MAC} units~\cite{Chaurasiya2018,Jaiswal2019,Murillo2021Energy,Zhang2019Efficient,Zhang2020Design,Uguen2019Evaluating,Crespo2022,murillo2023Generating} but just few designs have been directed towards the implementation of the division and/or the square root operations~\cite{Jaiswal2019,Rao2021Posit,Raveendran2020ANovel,Xiao2020Posit,murillo2023Generating,murillo2024Square,murillo2023Suite}.
These existing works have explored basic posit division, but there is potential to further optimize the hardware implementation of such an operation.

Division plays a crucial role in numerous algorithms used in essential applications, such as digital signal processing. However, its hardware implementations are characterized by longer latency and higher costs compared to the other three fundamental arithmetic operations, prompting designers to avoid using it whenever possible. Division is the most intricate of the four fundamental arithmetic operations. Unlike the other operations, it typically results in both a quotient and a remainder. In floating-point and posit arithmetic, the remainder is often ignored, and division yields an approximate value due to the possibility of the dividend not being a multiple of the divisor. To simplify the process, fast division algorithms are frequently employed~\cite{EL04}.

In posit arithmetic, division operates as a fixed-point division and requires extra hardware for managing exponents, which results in more complex divider circuits. 

The primary division algorithms rely on either digit-recurrence (which has linear convergence) or multiplicative methods (which have quadratic convergence). In~\cite{Nannarelli2016}, the energy efficiency of both methods is evaluated, demonstrating that digit-recurrence methods are significantly more energy-efficient and require less area than multiplicative methods. Due to these characteristics, this division method has recently attracted interest from researchers in the industry, as evidenced by the works in  \cite{Brug_div_2018,Brug_div_2020,Brug_div_2022,Brug_div_2023}. To be precise, the author of~\cite{Brug_div_2018} shows that, for half, single, and double precision, the proposed high-radix digit-recurrence algorithm achieves lower latency compared to previous digit-recurrence and multiplicative methods.

The main contribution of this work is the development of digit-recurrence posit division algorithms that advance current methodologies through targeted hardware optimizations. Our designs integrate redundant arithmetic representations, on-the-fly quotient conversion, and operand scaling, which together simplify the digit-recurrence process and substantially improve performance metrics. Unlike previous approaches, our proposal achieves a more efficient balance between hardware complexity and operational speed, resulting in significant reductions in latency, area, and power consumption. We validate our proposal with an extensive comparative evaluation of different design variants, including both radix-2 and radix-4 implementations for double, single and half-precision, synthesized using a 28~nm TSMC standard library. 
The proposed operators have exhibited substantial reductions in hardware resources, in addition to demonstrating higher performance than existing solutions.
These contributions not only provide a more competitive solution for posit division but also offer valuable insights into high-performance arithmetic unit design, thereby positioning our work as a critical advancement in the field.

The rest of the paper is organized as follows.
\cref{sec:background} describes some basic concepts when dealing with the posit arithmetic format and introduces the main posit dividers in literature. In \cref{sec:posit division algorithms}, the fundamentals of digit-recurrence division are reviewed, and the proposed digit-recurrence posit division algorithm is explained, along with the different optimizations.
Last, \cref{sec:hardware resources} presents the hardware evaluation of the implementation, and \cref{sec:conclusions} draws our final remarks and conclusions.

\IEEEpubidadjcol 

\section{Background}
\label{sec:background}

In this section we introduce the basics about posit arithmetic, and cover the literature about posit division.

\subsection{Posit Arithmetic}
\label{subsec:posit arithmetic}

A posit number configuration is typically specified by two parameters, $\langle n, es \rangle$: the total bit width $n$ and the exponent size $es$, which represents the number of bits allocated to the exponent field. While the most commonly used posit formats in the literature~\cite{Gustafson2017Beating, DeDinechin2019} have been \positenv{8}{0}, \positenv{16}{1}, and \positenv{32}{2}, the latest version of the Posit Standard~\cite{positworkinggroup2022Posit} standardizes the value of $es$ to 2. This fixed value simplifies hardware design and facilitates conversions between different posit sizes~\cite{Guntoro2020}. Consequently, this configuration will be adopted throughout the remainder of this paper, represented as \positbits{n}, where $n$ denotes the total bit width, and the exponent field is consistently 2 bits in length.

\begin{figure}[!t]
    \centering
    \includegraphics[width=1.0\linewidth]{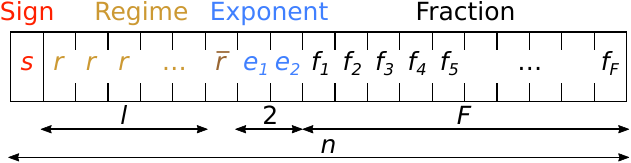}
    \caption{Generic $n$-bit posit binary encoding.}
    \label{fig:posit_format}
\end{figure}

Posit arithmetic defines only two special cases: zero and \ac{NaR}, represented by the bit patterns \texttt{0$\dots$0} and \texttt{10$\dots$0}, respectively. All other bit patterns are used to encode real values, consisting of four fields as illustrated in \cref{fig:posit_format}: a sign bit ($s$), a series of bits for the regime value ($k$), up to $es=2$ bits for the exponent ($e$), and the remaining bits for the normalized fraction ($f$).

The regime is a sequence of $l$ identical bits ($r$) finished with a negated bit ($\bar{r}$) that encodes an extra scaling factor $k$ given by \cref{eq:regime_value},
\begin{equation} \label{eq:regime_value}
    k = \left\{
	\begin{array}{ll}
		-l & \mbox{if } r_0 = 0 \\
		l-1 & \mbox{if } r_0 = 1
	\end{array}
	\right..
\end{equation}
As this field does not have a fixed length, it may cause the exponent to be encoded with less than 2 bits, even with no bits if the regime is wide enough. The same occurs with the fraction, which must be normalized with respect to the size of the fraction field ($2^F$).

Since the regime field does not have a fixed length, the number of bits available for encoding the exponent can be reduced, potentially leaving it with less than 2 bits, or even none, if the regime is sufficiently wide. A similar situation arises with the fraction, which must be normalized on the basis of the bit width of the fraction field ($2^F$). Therefore, if we consider the bit string for a posit $p$ as a signed integer, ranging from $-2^{n-1}$ to $2^{n-1}-1$, then $p$ represents the value given by \cref{eq:posit_value}:
\begin{equation} \label{eq:posit_value}
    X = \begin{cases}
        0  &   \text{if }  p =0  \\
        \text{NaR}  &  \text{if } p =-2^{n-1} \\
        (-1)^{s} 2^{4k + e}(1 + f)  &    \text{otherwise}
    \end{cases}~.
\end{equation}


The primary differences with respect to the IEEE 754 floating-point format are the existence of the regime field, the use of an unbiased exponent, and the interpretation of the hidden bit in the fraction. In traditional floating-point arithmetic, the hidden bit is 1 but for the subnormal numbers, for which is 0. However, for posits, the hidden bit is always 1; there are no subnormal numbers in posit arithmetic. Also, negative posits are represented in two's complement~\cite{Murillo2020Customized}.

Another difference is the fact that \ac{NaR} has a unique representation corresponding to the most negative two's complement signed integer. As a result, it is considered less than all other posit values and equal only to itself in comparison operations. Additionally, the remaining posit values maintain an order consistent with their bit representations. This property enables posit numbers to be compared directly as two's complement signed integers, removing the need for additional hardware dedicated to posit comparison operations~\cite{Mallasen2022}.

Posits also feature support for fused operations, but in a wider sense than the traditional floating-point \ac{MAC} operations. Posit fused operations involve more than two operands, and utilize a larger register known as the \emph{quire} to accumulate intermediate results. By avoiding intermediate roundings, fused operations deliver significantly higher accuracy~\cite{Murillo2021Energy, Crespo2022}.

All in all, despite the fact that posit arithmetic was designed to resemble floating-point circuitry, the variable-length fields and the signed hidden bit in the fraction necessitate some redesign of the logic for implementing posit operators. However, this additional effort can be justified by the advantages that posit arithmetic provides. Its higher accuracy compared to standard floating-point allows for reducing the bit width of data and operations without compromising result precision, bringing significant hardware-level benefits~\cite{murillo2023Generating,Wang2022PL-NPU,mallasen2025phee}.


\subsection{Related Work}
\label{subsec:related work}

PACoGen~\cite{Jaiswal2019} is a tool that generates \ac{HDL} code for several posit operations. Arguably it is the first work to include an open-source posit divider design, which features the Newton-Raphson algorithm. Additionally, works such as~\cite{Raveendran2020ANovel, Xiao2020Posit} introduce functional units for both division and square root operations, which share similar underlying digit-recurrence algorithms. The unit in~\cite{Raveendran2020ANovel} combines division and square root functionalities in a single hardware design, employing an iterative approach based on the non-restoring algorithm. In contrast, the designs in~\cite{Xiao2020Posit} apply the non-restoring method too. However, neither of these previous works are open-source. The authors of~\cite{murillo2023Suite} evaluate various implementations of posit division, exploring both iterative and multiplicative methods that enhance previous posit designs in terms of latency and resource efficiency. Literature also shows several complete \acp{PAU} that have been integrated in RISC-V cores, like PERCIVAL~\cite{mallasen2024BigPERCIVAL} or CLARINET~\cite{Sharma2023CLARINET} or approximate division and square root units like the work in~\cite{murillo2023PLAUs}, which applies the logarithmic transform~\cite{Mitchell1962} to the posit fractions.

\section{Digit-Recurrence Posit Division Algorithm}
\label{sec:posit division algorithms}

Let $X$ and $D$ be the dividend and divisor of a division in which both operands are posit numbers. Considering $p_X$ as the bit string of $X$, we have
\begin{equation} \label{eq:posit_value_new_X}
    X = \begin{cases}
        0  &   \text{if }  p_X =0  \\
        \text{NaR}  &  \text{if } p_X =-2^{n-1} \\
        (-1)^{s_X} 2^{4k_X + e_X}(1 + f_X)  &    \text{otherwise}
    \end{cases}~,
\end{equation}
and considering $p_D$ as the bit string of $D$, we have
\begin{equation} \label{eq:posit_value_new_D}
    D = \begin{cases}
        0  &   \text{if }  p_D =0  \\
        \text{NaR}  &  \text{if } p_D =-2^{n-1} \\
        (-1)^{s_D} 2^{4k_D + e_D}(1 + f_D)  &    \text{otherwise}
    \end{cases}~,
\end{equation}
The result of the operation,
\begin{equation}
Q = \frac{X}{D}~, \label{eq_X:D}
\end{equation}
is a $n$-bit posit number $Q$ and considering $p_Q$ as its bit string, we have
\begin{equation} \label{eq:posit_value_new_Q}
    Q = 
    \begin{cases}
        0  &   \text{if }  p_Q =0  \\
        \text{NaR}  &  \text{if } p_Q =-2^{n-1} \\
        (-1)^{s_Q} 2^{4k_Q + e_Q}(1 + f_Q)  &    \text{otherwise}
    \end{cases}~.
\end{equation}
If the divisor $p_D$ is either 0 or \ac{NaR}, or if the dividend $p_X$ is \ac{NaR}, the result is a \ac{NaR}. On the other hand, if just the dividend $p_X$ is 0, the result is also 0.
Let us focus on the case where both operands are different of 0 and \ac{NaR}, and let us call $x$ and $d$ to the fractional significands of the operands%
\footnote{The algorithms presented in this paper work for fractional operands in $[0.5,1)$. Although the posit format uses significands in the range $[1,2)$, the adaptation to this format is straightforward, and the calculations are equivalent.}
$X$ and $D$, respectively, that is $x = (1 + f_X)/2 $ and $d = (1 + f_D)/2$.
The major steps for a posit division are:
\begin{itemize}
    \item Extract the sign of the result ($s_Q$) from the two sign bits ($s_X , s_D)$, that is $s_Q = s_X \oplus s_D$.
    \item Subtract the scaling factors (exponents and regimes), and find a new pair of regime $k_Q$ and exponent $e_Q$ such that
    \begin{equation}
        4 k_Q  + e_Q= 4(k_X - k_D) + e_X - e_D = T.
    \end{equation}
    Since the posit exponent is a 2-bit unsigned integer, we have $0~\le~e_Q~<~4$, and it follows that
    \begin{gather}
        e_Q = T \bmod 4, \\
        k_Q = \left \lfloor \frac{T}{4} \right \rfloor.
    \end{gather}
    Notice that this can be easily computed in hardware if considering each of the scale factors $4k + e$ as signed numbers, so the two LSB of the result correspond to the exponent $e_Q$, and the rest to the regime value $k_Q$~\cite{murillo2023Suite}.
    \item Divide the significands $x$ and $d$:
    \begin{equation}
    q = \frac{x}{d}~, \label{eq_x:d}
    \end{equation}
    where $q$ is the quotient of the significands. It contains the significand of the quotient $Q$ (that is, $1+f_Q$) plus some extra bits for normalization and rounding.
    \item Normalize the result, if necessary, and decrement the exponent accordingly. Since posit significands lie in the range $[1, 2)$, its quotient $q$ falls in the range $[0.5, 2)$. Consequently, normalization is required when the quotient is less than $1$. In this case, a left shift by one position is required as well as a decrement of the exponent.
\end{itemize}

\begin{figure}[!t]
    \centering
    \includegraphics[width=1.0\linewidth]{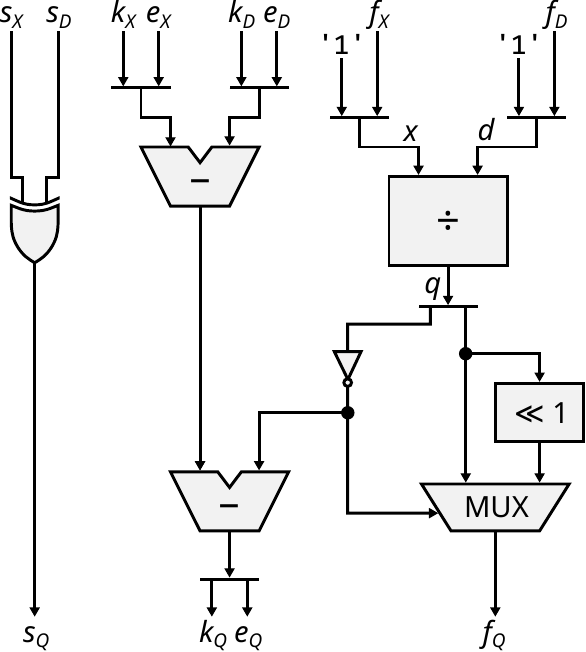}
    \caption{Basic scheme for posit division.}
    \label{fig:scheme}
\end{figure}

The block diagram from \cref{fig:scheme} illustrates the general architecture for posit division.
As aforementioned, the final regime $k_Q$ and exponent $e_Q$ can be obtained by assembling the regimes and exponents to subtract them together. The normalization process checks the \ac{MSB} of the quotient of the significands (that corresponds to the integer bit).
Notice that we are considering posits in a sign-magnitude notation, so the two's complement of negative inputs and outputs must be computed.

\subsection{Fractional division algorithm}

The digit recurrence algorithm for division consists of $It$ iterations of a recurrence, in which each iteration  produces one digit of the quotient~\cite{EL04} (the value of $It$ is discussed later in \cref{sec:ni}). This is preceded by an initialization step and followed by a termination step.

In what follows we show the digit recurrence step for division of posit significands (see~\cite{EL04} for a detailed description based on regular numbers). Let $q(i)$ denote the value of the quotient after $i$ iterations:
\begin{equation}\label{eq:quo_val}
q(i) = q(0) + \sum_{j=1}^i q_j r^{-j},
\end{equation}
where $q(0)$ is calculated in the initiation step, $r$ is the radix of the quotient and $q_j$ is the $j$-th digit of the quotient.
The quotient-digit set plays an important role in the characteristics of the algorithm. The canonical digit set $0 \leq q_j \leq r-1$ leads to the basic restoring division. On the other hand, using radix-2, the non-redundant digit set $\{-1, 1\}$ (no $0$ digit) results in a non-restoring algorithm, which is faster. The latter, depicted in \cref{alg:non_restoring}, has been employed in previous works implementing digit-recurrence posit division~\cite{Raveendran2020ANovel, Xiao2020Posit, murillo2023Suite} and is used as starting point for our evaluations, as well as for comparison purposes with previous implementations.

\begin{algorithm}[!t]
\caption{Non-restoring division}\label{alg:non_restoring}
\begin{algorithmic}[1]
    \Require{$0 \leq x < d$ of size $n$-bit}
    \Ensure{$x = d \times q + rem$}
    \State $w(0) \gets x$
    \Comment{Initialization}
    \For{$i = 0, \dots, n-1$}
    \Comment{Recurrence}
        \If{$w(i)\geq 0$}
            \State $q_{i+1} \gets 1$
        \Else
            \State $q_{i+1} \gets -1$
        \EndIf
        \State $w(i+1) \gets 2w(i) - dq_{i+1}$
    \EndFor
    \If{$w(n) < 0$}
    \Comment{Termination}
        \State $q \gets q(n) - ulp$
        \State $rem \gets w(n) + d$
    \Else
        \State $q \gets q(n)$
        \State $rem \gets w(n)$
    \EndIf
\end{algorithmic}
\end{algorithm}

\cref{alg:non_restoring} can be improved in two main ways. One approach is to use higher radices, which reduces the number of iterations. Another approach is to employ redundant arithmetic for updating the residual. However, due to the non-redundancy of the digit set, the quotient-digit selection becomes complex, as it requires the exact determination of the sign of the residual. The sign detection involves the conversion from redundant to conventional arithmetic of the residual in every iteration, which nullifies the benefit of using redundant arithmetic. A faster selection function can be achieved by using a \emph{redundant digit set} for the quotient. This allows the choice of a quotient-digit to be approximate, as any incorrect selection can be corrected in a later step. Such tolerance enables quotient-digits to be selected based on only a few of the \acp{MSB} of the shifted residual, rather than requiring a full-width subtraction.  
These issues are addressed in the following subsections, particularly in \cref{sec:oa,sec:sf}, where we discuss them in detail.
Such algorithms are known as \emph{SRT division}~\cite{Koren2018}.


In this work, we use a symmetric signed-digit set of consecutive integers for the quotient $q$ such as $q_i \in [-a,a]$, where $a \ge \lceil r/2 \rceil$ to keep a redundant representation. The redundancy factor is defined by \cref{eq:redundacy},
\begin{equation}\label{eq:redundacy}
\rho = \frac{a}{r-1},\ \ \ \ \frac{1}{2} < \rho \le 1.
\end{equation}
It is evident that in the case of radix-2, there is only one choice of $a=1$, and that $\rho$ is always equal to 1. Conversely, for higher radices there are multiple choices of the digit set. 
To illustrate this, consider the radix-4 case, in which two possibilities exist for the redundant digit set: $a=2$ and $a=3$ (maximum redundancy). While the former case has the advantage that the multiples of the divisor $dq$ are easier to generate, the case $a=3$ results in a simpler quotient-digit selection function.
In this work, for radix-4 division we consider $a=2$, so $\rho = 2/3$.

The digit recurrence algorithm is based on keeping a partial remainder (or residual) inside a convergence bound in each iteration. The residual $w(i)$ after $i$ iterations is defined as
\begin{equation}
w(i) = r^i (x-d q(i)),
\end{equation}
with bound to be kept
\begin{equation}
|w(i)| \le \rho d \label{eq:bound}.
\end{equation}
Then, from \cref{eq:quo_val} we can construct the  basic recurrence on which the division algorithms are based:
\begin{equation}\label{eq:convergence}
w(i+1) = r w(i)-dq_{i+1}~.
\end{equation}
The initial value of $w(0)$ depends on the redundancy factor $\rho$ and is determined during the initialization step (see \cref{sec:init_step}).
The recurrence is carried out in such a way that $w(i)$ is bounded by \cref{eq:bound}. The value of $q_{i+1}$ is selected according to the quotient-digit selection function, which is obtained as a function of a truncated version of $rw(i)$ and $d$ (see \cref{sec:sf}).

If the final residual is negative, a correction step is required. In such a case, the dividend has to be added to the final residual (if remainder is required), and the quotient is decremented by one \ac{ulp}. In posit arithmetic (as well as in floating-point), the sticky bit for proper rounding is provided by the remainder of the fraction division.

\begin{figure}[!t]
    \centering
    \includegraphics[width=1.0\linewidth]{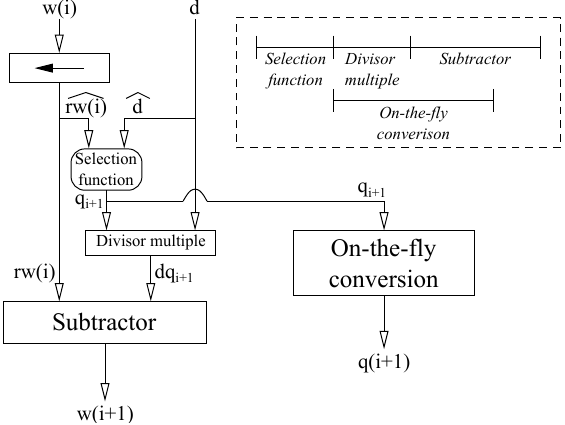}
    \caption{Basic modules and timing of division by digit recurrence.}
    \label{fig:digit_recurrence_architecture}
\end{figure}

All these steps are summarized in \cref{alg:srt} which is valid for a generic radix-$r$, while the basic modules and the timing of division by digit recurrence are shown in \cref{fig:digit_recurrence_architecture}.

\begin{algorithm}[!t]
\caption{Radix-$r$ digit recurrence division}\label{alg:srt}
\begin{algorithmic}[1]
    \Require{$0 \leq x < 1$, $1/2 \leq d < 1$, $x < d$ of size $n$-bit}
    \Ensure{$x = d \times q + rem$}
    \State $w(0) \gets x/p$
    \Comment{Initialization; $p=2$ if $\rho=1$, else $p=4$}
    \label{alg:line:initialization}
    \For{$i = 0, \dots, It-1$}
    \Comment{Recurrence}
        \State $q_{i+1} \gets SEL(rw(i), d) $
        \State $w(i+1) \gets rw(i) - dq_{i+1}$ \label{alg:line:subtraction}
    \EndFor
    \If{$w(It) < 0$}
    \Comment{Termination}
        \State $q \gets p(q(It) - ulp)$
        \State $rem \gets p(w(It) + d)$
    \Else
        \State $q \gets p\cdot q(It)$
        \State $rem \gets p\cdot w(It)$
    \EndIf
\end{algorithmic}
\end{algorithm}


\subsection{Optimizations of the algorithm}\label{sec:oa}

\subsubsection{Using redundant representation for the residual}

The residual $w(i)$ can be represented in non-redundant (i.e. conventional two's complement) or redundant form (carry-save or signed-digit). A first optimization consists in employing a redundant representation, since it results in a faster iteration (the subtraction $rw(i) - dq_{i+1}$ belongs to the critical path, see timing in \cref{fig:digit_recurrence_architecture}), although it increases the number of register bits required. 

\subsubsection{Fast sign and zero detection of the final residual}
A sign detection of the last residual is needed to carry out a possible correction, and the zero-remainder condition may also be required (depending on the application). In a redundant representation of the residual (i.e a carry-save implementation $w = wc+ws$) both the zero condition and the sign of the last residual are difficult to detect, since they involve a conversion from redundant to conventional representation. To solve this problem, a second optimization can be considered: a sign and zero detection lookahead network for the carry-save representation of the residual. This is proposed in~\cite{EL04}, which avoids the slow conversion.

\subsubsection{On-the-fly conversion}\label{sec:otfcr}

Another important optimization is the on-the-fly conversion of the quotient from signed-digit to conventional representation by overlapping it with the division iteration (at the cost of a hardware increase). This technique not only handles the conversion but it also keeps the required decremented value of the quotient when the final remainder is negative, simplifying the correction step.

In the following, we summarize the algorithm proposed in~\cite{EL04} which is adapted here to the posit quotient.
Let $Q(i)$ denote the digit vector of the converted quotient consisting of the $i$ most significant digits
\begin{equation}
Q(i) = \sum_{j=1}^i q_ir^{-i}~.
\end{equation}
To avoid carry propagation, a new form $QD(i)$ (decremented form) is defined as
\begin{equation}
QD(i) = Q(i) - r^{-i}~.
\end{equation}

The on-the-fly conversion algorithm is based on the concatenation of digits, as opposed to their addition, thus preventing the carry propagation.
The algorithm is derived from \cref{eq:Q,eq:QD} in terms of concatenations as follows:
\begin{gather}
    \scalebox{0.89}{$
    Q(i+1) = 
    \begin{cases}
        Q(i)\parallel q_{i+1}  &   \text{if }  q_{i+1} \ge 0  \\
        QD(i)\parallel (r-|q_{i+1}|)  &  \text{if } q_{i+1} < 0
    \end{cases}~,\label{eq:Q}
    $}
    \\
    \scalebox{0.89}{$ 
    QD(i+1) = 
    \begin{cases}
        Q(i)\parallel (q_{i+1}-1)  &   \text{if }  q_{i+1} > 0  \\
        QD(i)\parallel ((r-1)-|q_{i+1}|)  &  \text{if } q_{i+1} \leq 0
    \end{cases}~,\label{eq:QD}
    $}
\end{gather}
where $\parallel$ means bit concatenation and $Q(0)=QD(0)=0$. In essence, this algorithm retains the binary value of $Q(i)$ and its decremented binary counterpart, $QD(i)$, from a specific iteration onward, where $QD(i) = Q(i)-1$.


After the regular iterations, these equations are used to select $Q(It)$ (if the final remainder is greater than or equal to zero) or $QD(It)$ (if the remainder is negative) as the quotient of the division before the normalization and rounding steps.


\subsubsection{Operands scaling}\label{sec:scaling}

To simplify the quotient-digit selection function it is possible to scale 
the divisor to a value close to 1, in such a way that the selection function depends not on the divisor $d$, but only on a small number of bits of the partial remainder~\cite{EL_TC90,EL94}. 
It has been determined that, for a radix-4 digit
recurrence with digit set $\left\{-2, ..., 2\right\}$, it is enough to have the scaled divisor in the range $[1-1/64, 1+1/8]$~\cite{EL_TC90,EL94}.

The divisor is multiplied by a scaling factor $M$ which depends only on the divisor $d$ such that $M \times d \approx 1$. As shown in \cref{tab:scale_factor}, only three fractional bits of the divisor are needed to select the scale factor. Note that for the scaling, the divisor is supposed to be in the range $[0.5, 1)$. The multiplication by $M$ is achieved by a shift-add procedure (instead of using a regular multiplier), since the factor $M$ is decomposed as the addition of the divisor (component 1 in \cref{tab:scale_factor}) plus one or two multiplies of the divisor (by powers of 2).
The dividend has to be scaled by the same amount $M$ to get the correct result.
Additional bits are also required to allow the shifted dividend and divisor to be accurately computed.

\begin{table}
\centering
\ra{1.2}
\caption{Scaling factor ($M$) and the corresponding components for radix-4, $a=2$}\label{tab:scale_factor}
    \begin{tabular}{cccccc}
        \toprule
        $d$ & $M$  & Components\\
        \midrule 
        0.1\underline{000}xxx... & 2     & 1 + 1/2 + 1/2 \\
        0.1\underline{001}xxx... & 1.75  & 1 + 1/4 + 1/2 \\
        0.1\underline{010}xxx... & 1.625 & 1 + 1/2 + 1/8 \\
        0.1\underline{011}xxx... & 1.5   & 1 + 1/2       \\
        0.1\underline{100}xxx... & 1.375 & 1 + 1/4 + 1/8 \\
        0.1\underline{101}xxx... & 1.25  & 1 + 1/4       \\
        0.1\underline{110}xxx... & 1.125 & 1 + 1/8       \\
        0.1\underline{111}xxx... & 1.125 & 1 + 1/8       \\
        \bottomrule
    \end{tabular}
\end{table}

\subsection{Initialization process}\label{sec:init_step}

We have to take the fractional part of both posit operands to form the significands. In this paper, we consider fractional division algorithms for non-negative operands.
Although posit numbers can be processed efficiently with negative fractions~\cite{Murillo2022Comparing}, this is not the case for the division operation, as the introduction of an additional sign bit further complicates the application of the digit-recurrence algorithm~\cite{Koren2018} as well as the final normalization~\cite{murillo2023Suite}.
Therefore, the first step is to obtain the two's complement of the operands if they are negative.
On the other hand, the length of the fractional part of a Posit$n$ number is not fixed, unlike with conventional floating-point numbers. We have to consider the worst case, that is, the case in which the fraction has the maximum number of bits, that is $n-5$ bits.

The initial value $w(0)$ of the recurrence \cref{eq:convergence} has to fulfill the bound given by \cref{eq:bound}, that is $|w(0)| \le \rho d$. In what follows, we assume that $q(0)=0$, since it involves less hardware cost~\cite{EL04}. Let us study the case of $\rho = 1$ and  $1/2~<~\rho~<~1$ separately.
\begin{itemize}
    \item $\rho = 1$. 
        Since a posit significand is in the range $[1,2)$, it fulfills that $1 \leq d < 2$ and then
        \begin{equation}\label{eq:rd}
        1 \leq \rho d~.
        \end{equation}
         On the other hand, we select $w(0)=x/2$. Similarly we have, $1 \leq x < 2$, and dividing by 2 we have:
        \begin{equation}\label{eq:eqw2}
        \frac{1}{2}  \leq w(0) < 1 ~.
        \end{equation}
        From \cref{eq:rd,eq:eqw2} we conclude that
        \begin{equation}\label{eq:kk10}
        |w(0)| < \rho d~.
        \end{equation}
    \item $\frac{1}{2} < \rho < 1$. 
        As in the previous case, $1 \leq d < 2$, so
        \begin{equation}\label{eq:kk1}
        \frac{1}{2} \leq \rho d~.
        \end{equation}
         On the other hand, we select $w(0)=x/4$. Since $1 \leq x < 2$, dividing by 4 we have:
        \begin{equation}
        \frac{1}{4} \leq w(0) < \frac{1}{2}~. \label{eq:kk2}
        \end{equation}
        From \cref{eq:kk1,eq:kk2} we conclude that
        \begin{equation}\label{eq:kk11}
        |w(0)| \leq \rho d~.
        \end{equation}
\end{itemize}
As consequence, to initialize $w$ the value of  $x$ has to be right shifted either one position ($w(0)=x/2$, if the selected digit set has maximum redundancy) or two positions ($w(0)=x/4$, for a digit set without maximum redundancy).



\subsection{Selection function}\label{sec:sf}

The quotient-digit selection function depends on the radix, as well as whether the residual is represented using redundant arithmetic or not.

\subsubsection{Radix-2 with non-redundant residual}


For radix-2, i.e. with quotient-digit set $\{-1, 0, 1\}$, the quotient-digit selection function $SEL$ from \cref{alg:srt} does not depend on the divisor, just on the truncated shifted residual, namely $\widehat{2w(i)}$. More precisely, if the latter is represented in non-redundant notation, the selection function is defined by \cref{eq:r2_srt_nr_sel}
\begin{equation}\label{eq:r2_srt_nr_sel} 
    SEL(\widehat{2w(i)}) = \begin{cases}
                    +1 & \text{if } 1/2 \leq \widehat{2w(i)}\\
                    ~~0 & \text{if } -1/2 \leq \widehat{2w(i)} < 1/2 \\ 
                    -1 & \text{if } \widehat{2w(i)} < -1/2
        \end{cases}~.
\end{equation}
Since it only requires the comparison with the constants $-1/2 = (1.0)_2$ and $1/2 = (0.1)_2$, only the two \acp{MSB} of the shifted residual are required for the selection rules.


To further speed up the recurrence stage of the algorithm, we can also improve the partial remainder computation, so that an $n$-bit subtraction is not required in each step. This can be done by keeping the partial remainder in redundant form, like carry-save notation, so that $w = w_{s} + w_{c}$, as shown below.

\subsubsection{Radix-2 with redundant residual}
For radix-2 with the partial remainders in carry-save form, 
the subtraction in the recurrence (line~\ref{alg:line:subtraction} of \cref{alg:srt}) can be performed as a carry-save addition. To that end, we initialize $w_{s}(0) = x/2$ or $x/4$ and $w_{c}(0) = 0$ (line~\ref{alg:line:initialization} of \cref{alg:srt}). Also the selection function changes in this case; the quotient-digit selection function is now depicted by \cref{eq:r2r} (see~\cite{EL04}):
\begin{equation}\label{eq:r2r} 
    SEL(\widehat{2w(i)}) = \begin{cases}
                    +1 & \text{if } 0 \leq \widehat{2w(i)} \leq 3/2\\
                     ~~0 & \text{if } \widehat{2w(i)} = -1/2 \\ 
                    -1 & \text{if } -5/2 \leq \widehat{2w(i)} \leq -1
        \end{cases}~,
\end{equation}
with $\widehat{2w(i)}$ being the value of the carry-save shifted residual (that is $2(w_s(i)+2w_c(i))$) truncated to the fourth \ac{MSB} (three integer bits and one fractional bit), as suggested in~\cite{EL04}. Moreover, it has been empirically established that residual estimate never reaches the upper bound $3/2$, so just three bits (two integer, one fractional) from the carry-save shifted residual are good enough to make a proper selection of the quotient-digit~\cite{Sutter2004Power}.

\subsubsection{Radix-4 with redundant residual}
For radix-4 with redundant (carry-save) residual, the selection of the quotient-digit depends not just on the truncated carry-save shifted residual $\widehat{4w(i)}$ (as in the radix-2 case), but also on the truncated divisor $\hat{d}$. 
This selection is described in terms of some constants $m_k(\hat{d})$, as indicated by \cref{eq:r4_sel},
\begin{equation}\label{eq:r4_sel}
    SEL(\widehat{4w(i)}, \hat{d}) = k \text{~~ if } m_k(\hat{d}) \leq \hat{d} < m_{k+1}(\hat{d}),
\end{equation}
where $k \in [- a, a]$ and the selection bounds $m_k(\hat{d})$ are chosen as described in~\cite{EL04}. In this case, we consider the minimally redundant digit set $\{-2, -1, 0, 1, 2\}$, so for the selection function from \cref{eq:r4_sel} the divisor is truncated to four bits, and the shifted residual is truncated to the fourth fractional bit, with three integer bits for a total of seven bits.

\subsubsection{Radix-4 with operand scaling}

As discussed in \cref{sec:scaling}, it is possible to reduce the complexity of the selection function by scaling the divisor (and dividend) to a value close to 1, so that the selection function is independent of the divisor.
In this case, for radix-4, $a=2$ and redundant (carry-save) residual, the quotient-digit selection function is indicated by \cref{eq:r4r} (see~\cite{EL04}):
\begin{equation}\label{eq:r4r} 
    SEL(\widehat{4w(i)}) = \begin{cases}
        +2  & \text{if } ~~~~~3/2 \leq \widehat{4w(i)} \leq 3    \\
        +1  & \text{if } ~~~~~1/2 \leq \widehat{4w(i)} \leq 11/8 \\
        ~~0 & \text{if } ~-1/2    \leq \widehat{4w(i)} \leq 3/8  \\
        -1  & \text{if } -13/8    \leq \widehat{4w(i)} \leq -5/8 \\
        -2  & \text{if } -13/4    \leq \widehat{4w(i)} \leq -7/4 \\
    \end{cases}~,
\end{equation}
where $\widehat{4w(i)}$ corresponds to the conversion from redundant to conventional of the 6 \acp{MSB} of $w(i)$ (that is, $w_s(i)+2w_c(i)$ truncated to 6 bits).

\subsection{Data path bitwidth and iterations}\label{sec:nidpw}

\subsubsection{Number of bits of the residue data path}

The data path for residue computation consists of the number of bits of the posit significand (worst case $n-5$ fractional bits plus the integer bit, that is $n-4$ bits), plus one or two additional bits due to initial scaling (depending on the redundancy factor $\rho$, given by the function $(2 - \lfloor \rho \rfloor)$), plus the $log_2 r$ bits due to the shift involved in $r \cdot w(i)$ (see \cref{eq:convergence}). The total number of bits required for this computation is thus $n~-~2~+~log_2 r~-~\lfloor \rho \rfloor$.

\subsubsection{Number of iterations}\label{sec:ni}


Let $h$ denote the number of bits of the result needed to obtain the quotient with the required precision. It consists of the number of bits of the significand plus some extra bits, namely:

\begin{itemize}
\item One guard bit due to normalization (bit $G$). Since $q~\in~[0.5, 2)$, when $q<1$ a shift of one position to the left is required for normalization.
\item One rounding bit (bit $R$), which is necessary when the result is not normalized (posit arithmetic uses round to nearest).
\item One bit if $\rho = 1$ or two bits if $\rho \in \left(\frac{1}{2}, 1\right)$ (scaling due to initialization step, see \cref{sec:init_step}). However, notice that this initialization would make $w(0) < d$, and the quotient (which should be shifted to adjust for this initialization) will always contain a 0 bit in the \ac{MSB} ($q(0) = 0$), so actually such a bit can be omitted when implementing in hardware.
This can be denoted by the function $(1 - \lfloor \rho \rfloor)$.
\end{itemize}

Thus, the number of bits of the result required to calculate the quotient ($h$) is
\begin{equation}
h = n - 4 + 2 + (1 - \lfloor \rho \rfloor) = n - 1 -\lfloor \rho \rfloor~. \label{eq:w}
\end{equation}
For example, for radix-4 and digit set $\{ \pm 2, \pm 1, 0 \}$ ($\rho = 2/3$) the value of $h$ is $h = n-1$ bits.


In digit-recurrence division, the number of iterations $It$ depends on the number of bits required for calculating the quotient $(h)$ and the radix, as shown by \cref{eq:iters}:
\begin{equation}\label{eq:iters}
It = \left \lceil \frac{h}{\log_2 r} \right \rceil~.
\end{equation}

\subsubsection{Latency of the operators}
The total number of iterations of the different radices and data precisions are shown in \cref{tab:iters}. The number of clock cycles is also shown for pipelined operators. We set one cycle per iteration of the algorithm, plus three extra cycles: one for the termination step (see \cref{sec:termination_step}), and the other two for the posit decoding and encoding stages.
If using operand scaling, then one extra clock cycle is required for such a process.

\begin{table}
\centering
\ra{1.2}
\caption{Number of iterations in the digit iterations stage and latency of the division units.}\label{tab:iters}
    \begin{tabular}{lccccc}
    \toprule
         & & \multicolumn{2}{c}{Radix-2} & \multicolumn{2}{c}{Radix-4, $\rho < 1$} \\
     \cmidrule(lr){3-4} \cmidrule(lr){5-6}
         & \begin{tabular}[c]{@{}c@{}}Significand\\ bits\end{tabular} & Iterations & Latency & Iterations & Latency \\
    \midrule
    Posit16 & 12 & 14 & 17 & 8  & 11 \\
    Posit32 & 28 & 30 & 33 & 16 & 19 \\
    Posit64 & 60 & 62 & 65 & 32 & 35 \\
    \bottomrule
    \end{tabular}
\end{table}

\subsection{Termination process}\label{sec:termination_step}

In order to correctly produce the result, a termination step must be included. In this regard, the following actions must be performed:  
\begin{enumerate}  
\item Correction step: The recurrence may produce a negative final remainder. In such a case, the quotient must be corrected by subtracting one \ac{ulp} from the \ac{LSB} of the computed quotient, that is the bit $R$.

\item Compensation for the initial scaling step: Since there is an initial right shift of one or two positions, the quotient must be left-shifted accordingly.  


\item Normalization step: After the previous steps, the quotient will fall within the range $[0.5, 2)$. If it lies in $[0.5, 1)$, an additional left shift is required, so the quotient is normalized in the range $[1, 2)$, and thus the exponent must be decremented accordingly.

\item Rounding step: once the quotient is encoded in the posit format, we add 1 to the rounding bit position of the normalized fraction part. The sticky bit is given by the zero condition of the remainder. 
\end{enumerate}

The termination and rounding processes are illustrated in \cref{tab:round_example} for \positbits{10} format. The two examples present the same dividend $X$, while the divisor $D$ in the second example has one regime bit more (that is, divided by $2^4$). Obviously, the fractions in both examples are the same, and so are the computed quotients $q$ and remainders $rem$. The quotient of the fractions $q$ contains the integer bit plus two additional bits (guard and round). In this case, since the integer bit is \texttt{0}, that means $q < 1$, so it must be normalized, and the exponent decremented (so now $e_Q=1$). Thus, the resulting normalized fraction (without the integer bit) is $f_Q := \texttt{11110}$. The round bit follows the normalized fraction, and the sticky bit is provided by the ``remainder not equal to zero'' condition.

\begin{table}
\centering
\begin{threeparttable}[b]
\ra{1.2}
\caption{Termination and rounding examples for posit division}
\label{tab:round_example}
    \begin{tabular}{lcc}
    \toprule
              & Example 1           & Example 2     \\
    \midrule
    $X$       & \texttt{0011010111} & \texttt{0011010111} \\
    $D$       & \texttt{0001001100} & \texttt{0000100110} \\
    \midrule
    $k_Q$     & +1                  & +2         \\
    $e_Q$     & 2                   & 2          \\
    $q = x/d$ & \texttt{0.111110}\tnote{g}~\texttt{1}\tnote{r}  & \texttt{0.111110}\tnote{g}~\texttt{1}\tnote{r}  \\
    $rem$     & $\neq 0$            & $\neq 0$   \\
    $f_Q$     & \texttt{111101}\tnote{r}~\texttt{1}\tnote{s} & \texttt{111101}\tnote{r}~\texttt{1}\tnote{s} \\
    $f_{Q~shifted}$     & \texttt{11110}\tnote{r}~\texttt{1}\tnote{s} & \texttt{1111}\tnote{r}~\texttt{1}\tnote{s} \\
    $f_{Q~rounded}$     & \texttt{1111} & \texttt{000} \\
    \midrule
    $Q$       & \texttt{0110011111} & \texttt{0111010000} \\
    \bottomrule
    \end{tabular}
    \begin{tablenotes}
        \item [g] Guard bit.
        \item [r] Round bit.
        \item [s] Sticky bit.
    \end{tablenotes}
\end{threeparttable}
\end{table}

As demonstrated in these examples, the same quotients might result in different fractions when encoded in posit format, depending on the resulting regime. In the first example, the resulting fraction $f_Q$ is right-shifted by one bit due to the regime value $k_Q = 1$. Consequently, the final round bit is set to \texttt{0}, and no actual increment is performed. However, in the second example, the fraction is shifted two bits to the right due to the regime $k_Q = 2$. This results in a different fraction, and even the exponent is incremented due to the rounding carry.
Therefore, it is not possible to integrate normalization and rounding in the last digit iteration in the same manner to that employed in floating-point arithmetic~\cite{EL04}. 

\section{Hardware Evaluation}
\label{sec:hardware resources}

The proposed designs are evaluated in terms of area, power, timing, and energy (power-delay product) using Synopsys Design Compiler with a 28~nm TSMC standard library.
For every design, we consider both combinational and pipelined implementation. The pipelined designs compute one iteration per cycle, with two additional cycles for initialization and termination.
Recall that the number of iterations does not depend on the algorithm, but just on the radix (see \cref{sec:ni}).

\cref{tab:alg_sumamry} summarizes the different division algorithms and optimizations that have been implemented and evaluated upon different radices and formats.
The non-restoring division (NRD, \cref{alg:non_restoring}) has been incorporated into the evaluation process. This serves two primary functions: firstly, it establishes a baseline; secondly, it facilitates comparison with previous works implementing digit recurrence posit division~\cite{murillo2023Suite}.
The SRT division (see \cref{alg:srt}) is implemented with different optimizations: carry-save residual (CS), on-the-fly conversion (OF), and fast sign and zero detection of the final residual (FR). The SRT division with non-redundant residual is just implemented in radix-2, while both radix-2 and radix-4 are implemented with carry-save residual.
For the case of radix-4, operand scaling is also considered (see \cref{sec:scaling}), for which one extra cycle is needed.


\begin{table}
\centering
\ra{1.2}
\caption{Summary of implemented division algorithms}
\label{tab:alg_sumamry}
    \begin{tabular}{lcccccc}
    \toprule
    Implementation   & \begin{tabular}[c]{@{}c@{}}Redundant\\ residual\end{tabular} & \begin{tabular}[c]{@{}c@{}}On-the-fly\\ conversion\end{tabular} & \begin{tabular}[c]{@{}c@{}}Fast remainder\\ sign detection\end{tabular} & Radix     \\
    \midrule
    NRD          & \NO  & \NO  & \NO  & 2     \\
    SRT          & \NO  & \NO  & \NO  & 2      \\
    SRT CS       & \YES & \NO  & \NO  & 2 \& 4 \\
    SRT CS OF    & \YES & \YES & \NO  & 2 \& 4 \\
    SRT CS OF FR & \YES & \YES & \YES & 2 \& 4 \\
    \bottomrule
    \end{tabular}
\end{table}

\begin{figure}[!t]
    \centering
    \includegraphics[width=0.49\linewidth]{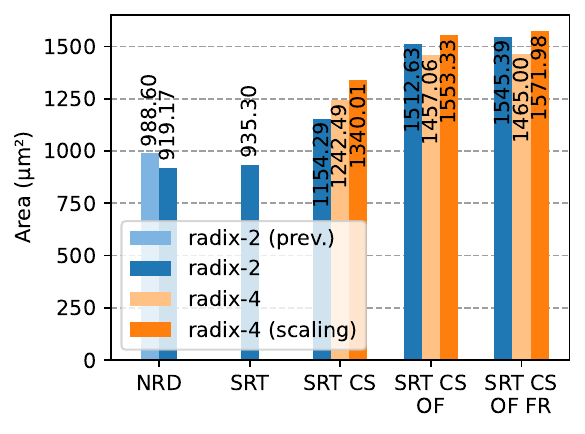}
    \includegraphics[width=0.49\linewidth]{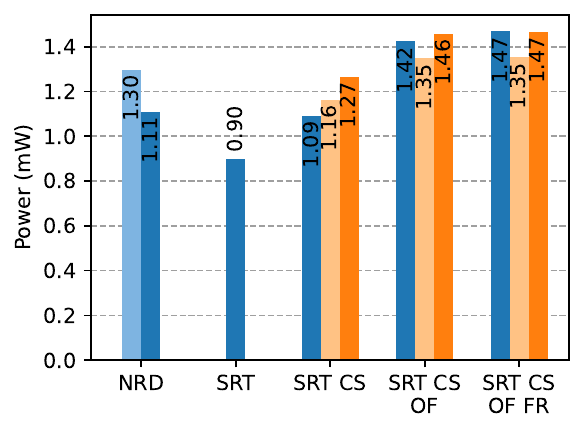} \\
    \includegraphics[width=0.49\linewidth]{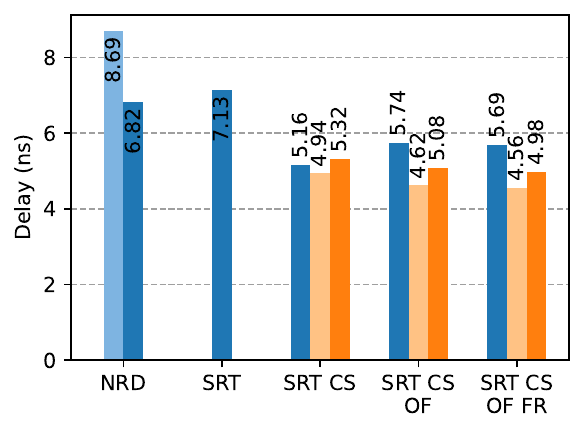}
    \includegraphics[width=0.49\linewidth]{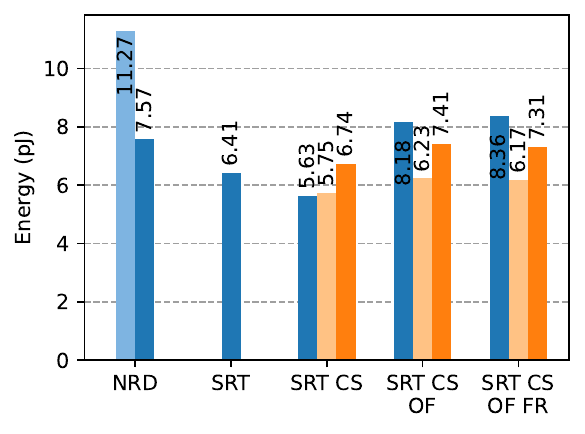}
    \caption{Synthesis results of combinational 16-bit posit dividers.}
    \label{fig:16_bit_comb}
\end{figure}

\begin{figure}[!t]
    \centering
    \includegraphics[width=0.49\linewidth]{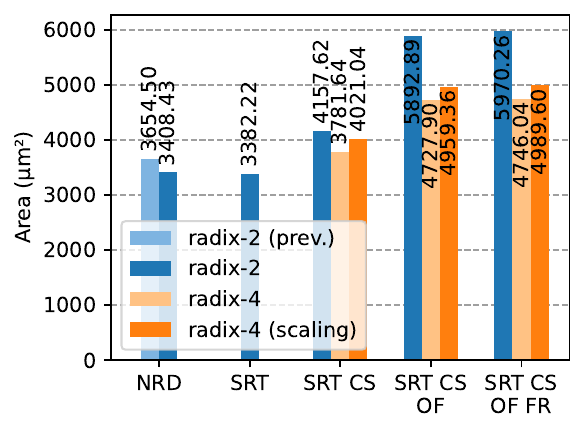}
    \includegraphics[width=0.49\linewidth]{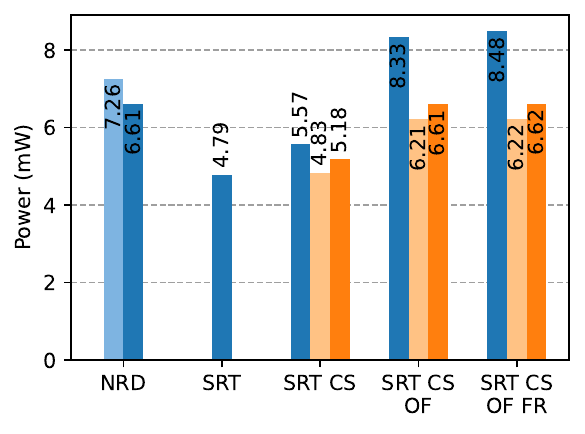} \\
    \includegraphics[width=0.49\linewidth]{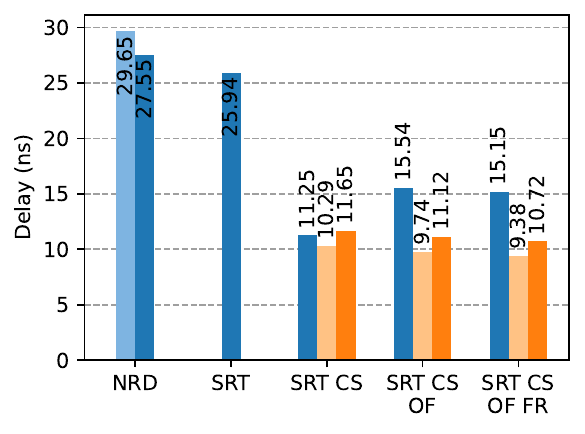}
    \includegraphics[width=0.49\linewidth]{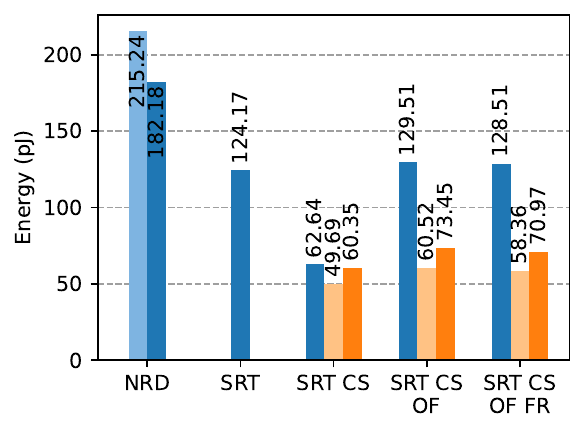}
    \caption{Synthesis results of combinational 32-bit posit dividers.}
    \label{fig:32_bit_comb}
\end{figure}

\begin{figure}[!t]
    \centering
    \includegraphics[width=0.49\linewidth]{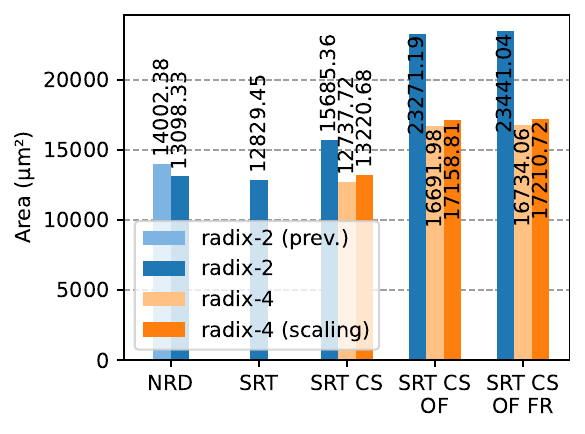}
    \includegraphics[width=0.49\linewidth]{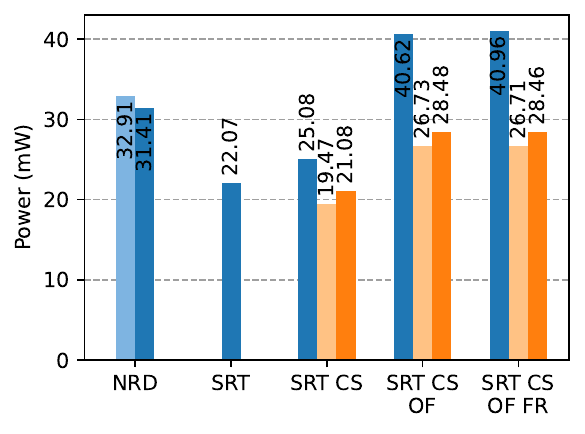} \\
    \includegraphics[width=0.49\linewidth]{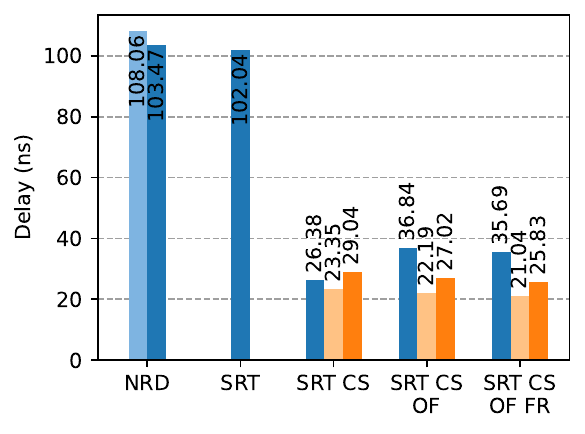}
    \includegraphics[width=0.49\linewidth]{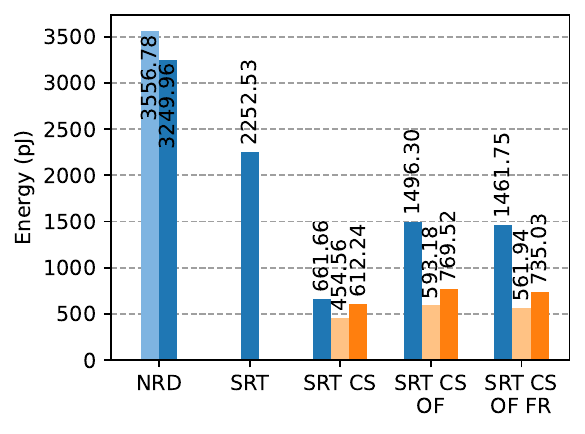}
    \caption{Synthesis results of combinational 64-bit posit dividers.}
    \label{fig:64_bit_comb}
\end{figure}

\cref{fig:16_bit_comb,fig:32_bit_comb,fig:64_bit_comb} show synthesis results for combinational dividers, with no timing restriction.
Similar results are obtained for the evaluated bitwidths, although more pronounced differences are obtained for larger datapaths.

The NRD and plain SRT radix-2 designs generally occupy the least area, confirming their simplicity in hardware implementation. The SRT optimized designs (CS, OF, FR) show a significant increase in area, especially when on-the-fly (OF) optimization is introduced. However, the radix-4 designs tend to occupy less area than radix-2, which is expected due to reduced number of iterations. 

In terms of delay, radix-4 implementations are superior to radix-2. Nevertheless, it must be noted that the most significant delay reduction is obtained in the CS variant, demonstrating the effectiveness of such optimization. 
On the other hand, introducing OF in radix-2 dividers slightly increases the delay, even though the termination step is faster with this optimization. 
In this particular scenario, the recurrence is so simple that it is faster than the on-the-fly update.
A similar problem occurs in the design of~\cite{Brug_div_2018}, where the author justifies the absence of on-the-fly conversion by stating that it increases the cycle time of the design.
The radix-4 with scaling variant does not significantly reduce the delay compared to plain radix-4, suggesting that scaling optimizations may not be effective in reducing the critical path combinational designs.

The power consumption is directly correlated to the area usage of each operator. Despite having higher power, radix-4 implementations tend to have competitive energy consumption due to their reduced delay. Overall, radix-4 dividers offer better speed and energy efficiency at the cost of higher area and power consumption, but such an overhead is amortized for larger datapaths, as depicted in \cref{fig:64_bit_comb}.

Compared with the previous work~\cite{murillo2023Suite}, the proposed NRD (radix-2) units present about 7\% less area, and delay reduced from 4.2\% (for \positbits{64}) up to 21.5\% (for \positbits{16}). The reason for such a noticeable reduction is that in~\cite{murillo2023Suite} posits are decoded in two's complement notation. This approach handles signed significands in $[-2, -1) \cup [1, 2)$, thereby necessitating an additional iteration of the digit-recurrence algorithm. The rest of posit dividers proposed in this paper demonstrate even larger reductions in delay and energy consumption compared to such previous work, with just a small area overhead.
For instance, the proposed radix-2 posit SRT CR dividers present a delay reduction of 40.6\% (for \positbits{16}), 62.1\% (for \positbits{32}), and 75.6\% (for \positbits{64}) with an area increment of just 16.8\%, 13.8\%, and 12\%, respectively. The energy is also decremented by 50.2\%, 70.9\% and 81.4\% for the same formats.
The proposed radix-4 posit SRT CR dividers present even higher resource savings for the 32-bit, and specially for the 64-bit scenario, were also the area is reduced by 9\% in comparison with the implementation from~\cite{murillo2023Suite}.

\begin{figure}[!t]
    \centering
    \includegraphics[width=0.49\linewidth]{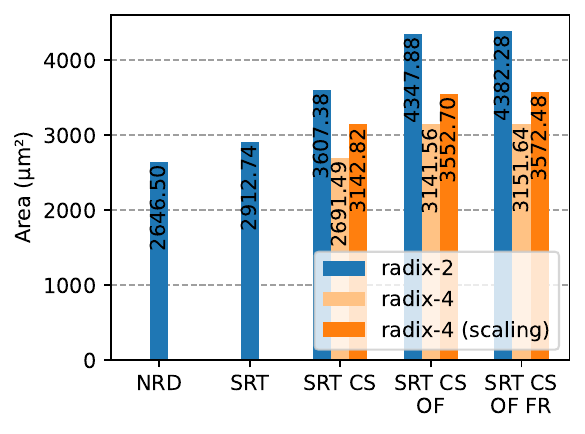}
    \includegraphics[width=0.49\linewidth]{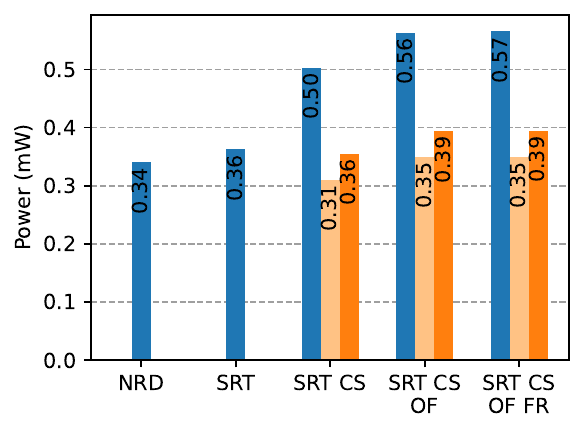} \\
    \includegraphics[width=0.49\linewidth]{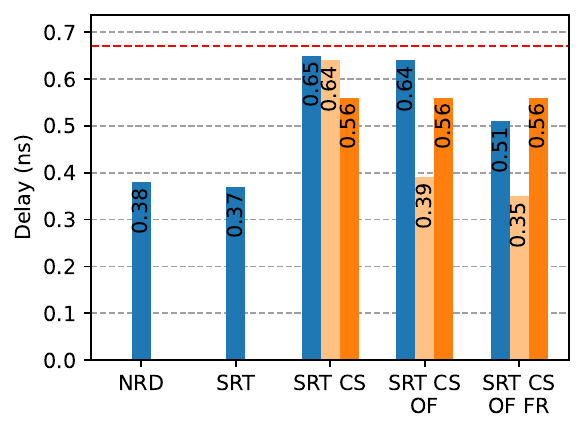}
    \includegraphics[width=0.49\linewidth]{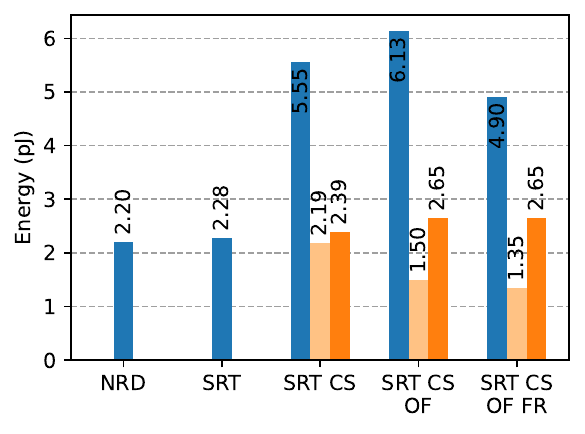}
    \caption{Synthesis results of pipelined 16-bit posit dividers.}
    \label{fig:16_bit_pipe}
\end{figure}

\begin{figure}[!t]
    \centering
    \includegraphics[width=0.49\linewidth]{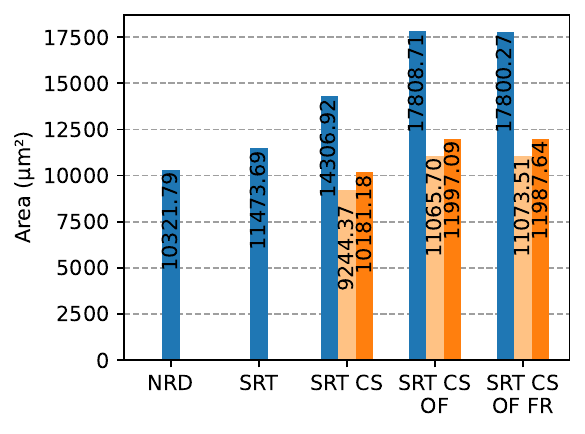}
    \includegraphics[width=0.49\linewidth]{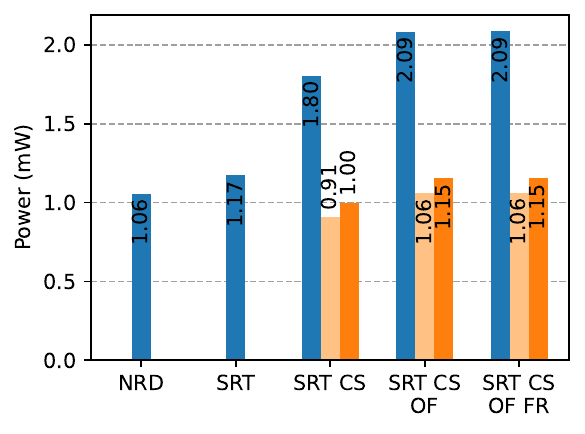} \\
    \includegraphics[width=0.49\linewidth]{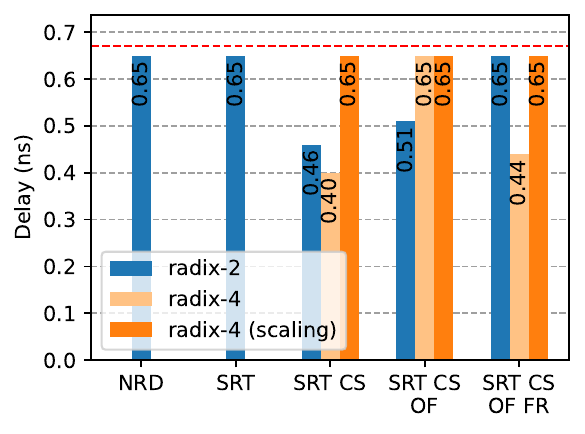}
    \includegraphics[width=0.49\linewidth]{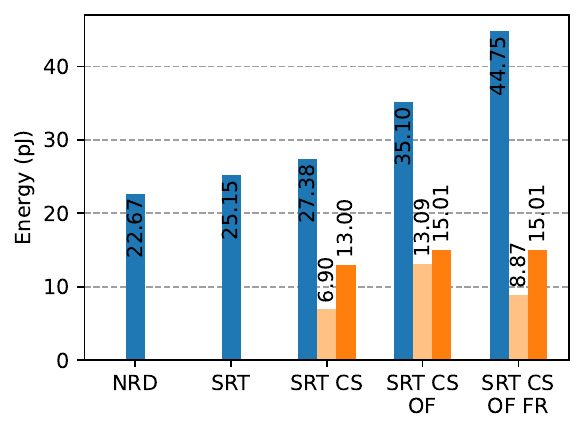}
    \caption{Synthesis results of pipelined 32-bit posit dividers.}
    \label{fig:32_bit_pipe}
\end{figure}

\begin{figure}[!t]
    \centering
    \includegraphics[width=0.49\linewidth]{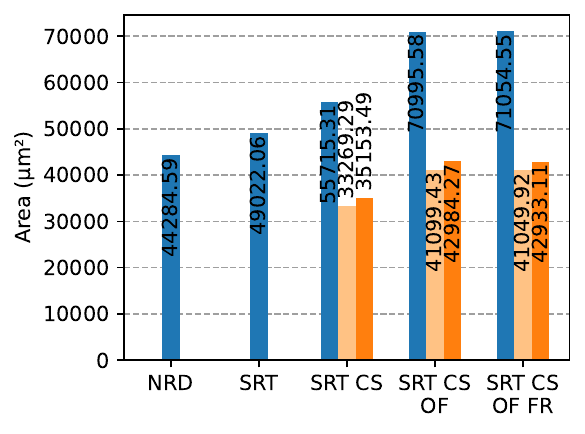}
    \includegraphics[width=0.49\linewidth]{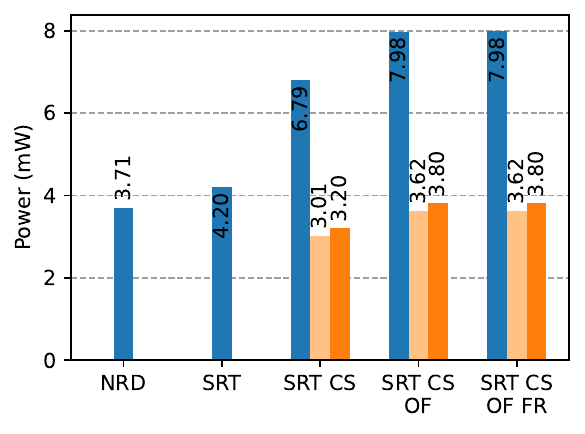} \\
    \includegraphics[width=0.49\linewidth]{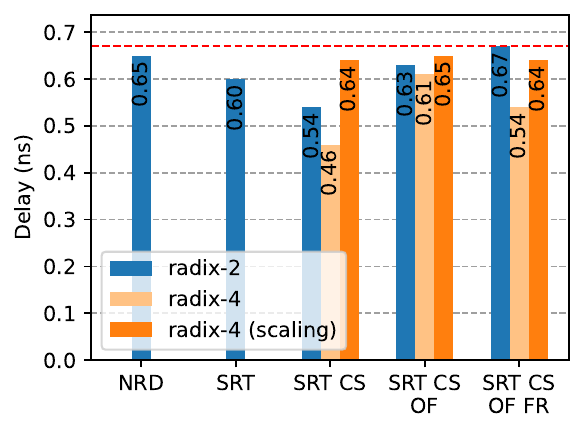}
    \includegraphics[width=0.49\linewidth]{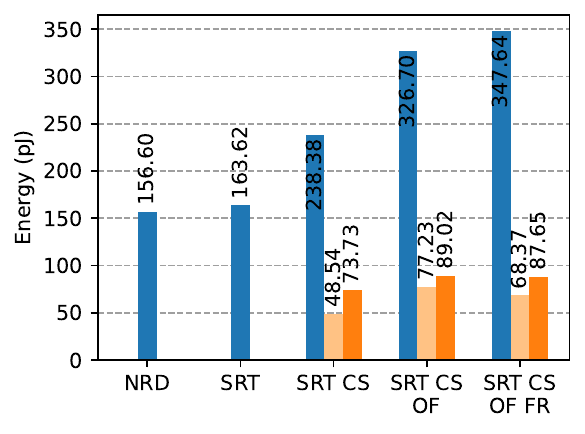}
    \caption{Synthesis results of pipelined 64-bit posit dividers.}
    \label{fig:64_bit_pipe}
\end{figure}

The results for pipelined operators are shown in \cref{fig:16_bit_pipe,fig:32_bit_pipe,fig:64_bit_pipe}. In this case, the target frequency has been set to 1.5~GHz.

As can be seen, although successive optimization techniques produce similar area and power increments as in the combinational scenario, there is a noticeable resource reduction when using radix-4 rather than radix-2. In this pipelined case, a lower number of iterations also has an impact on the amount of sequential elements like registers. This reduction in the number of cycles, in conjunction with the fact that all designs present a similar maximum delay (meeting the timing constraint), renders the radix-4 designs a significantly more energy-efficient solution.

In fact, the timing reports show that the critical path is not in the iterative stages, but in the final posit conversion and rounding phase (except for the designs with operand scaling, for which the longest path is in the scaling stage).
These observations underscore the fact that scaling the operands with the intention of simplifying the quotient-digit selection function does not result in any enhancement with regard to faster iterations. It is therefore recommended that a higher radix be employed in order to benefit from this optimization.

\section{Conclusions}
\label{sec:conclusions}

This work presented a comprehensive study on the adaptation and implementation of digit-recurrence division algorithms for posit arithmetic, introducing the first radix-4 implementation in this context. By leveraging classical digit-recurrence techniques and tailoring them to the posit number system, we provided efficient and scalable hardware designs that balance performance and resource consumption.

Our study considered both radix-2 and radix-4 iterations, each offering distinct trade-offs between area, power, and latency. The radix-2 implementations maintain a simpler hardware structure with lower area and power requirements, making them suitable for resource-constrained environments. However, they require a larger number of iterations, leading to increased division latency. On the other hand, radix-4 implementations significantly reduce the number of iterations and overall delay, resulting in faster computation at the cost of slightly increased area and power consumption. To further enhance efficiency, we incorporated optimizations such as redundant arithmetic for the residual, on-the-fly quotient conversion, and operand scaling, reducing the critical path and improving convergence properties.

We implemented and evaluated our designs across three different posit precisions: 16-bit, 32-bit, and 64-bit. The synthesis results using a 28~nm TSMC standard library revealed that radix-4 division consistently outperforms radix-2 in terms of speed, while still maintaining reasonable area and power overheads. Additionally, pipelined implementations demonstrated improved throughput, with radix-4 versions showing significant energy efficiency gains due to the reduced number of iterations.
Compared to existing posit division designs, including prior digit-recurrence approaches, our baseline, non-optimized implementations show improvements in hardware efficiency, execution time, and energy consumption, achieving up to a 27\% reduction in delay and approximately 7\% lower area usage.
More optimized designs demonstrate significantly greater resource reductions, with up to 81.4\% less power consumption at the cost of just 12\% more area.

These findings provide valuable insights into efficient posit-based arithmetic unit design, paving the way for future research on high-performance, low-power arithmetic operations in emerging computing architectures.




\section*{Acknowledgments}
%
%
This publication is part of the projects
PID2021-123041OB-I00, funded by MICIU/AEI/10.13039/501100011033 and by ERDF, EU, and
PID2022-136575OB-I00, funded by MICIU/AEI/10.13039/501100011033 and by ERDF, EU, and
PDC2023-145800-I00, funded by MICIU/AEI/10.13039/501100011033 and by European Union NextGenerationEU/PRTR.


\bibliographystyle{IEEEtran}
\bibliography{references,unificado}


 


\begin{IEEEbiography}[{\includegraphics[width=1in,height=1.25in,clip,keepaspectratio]{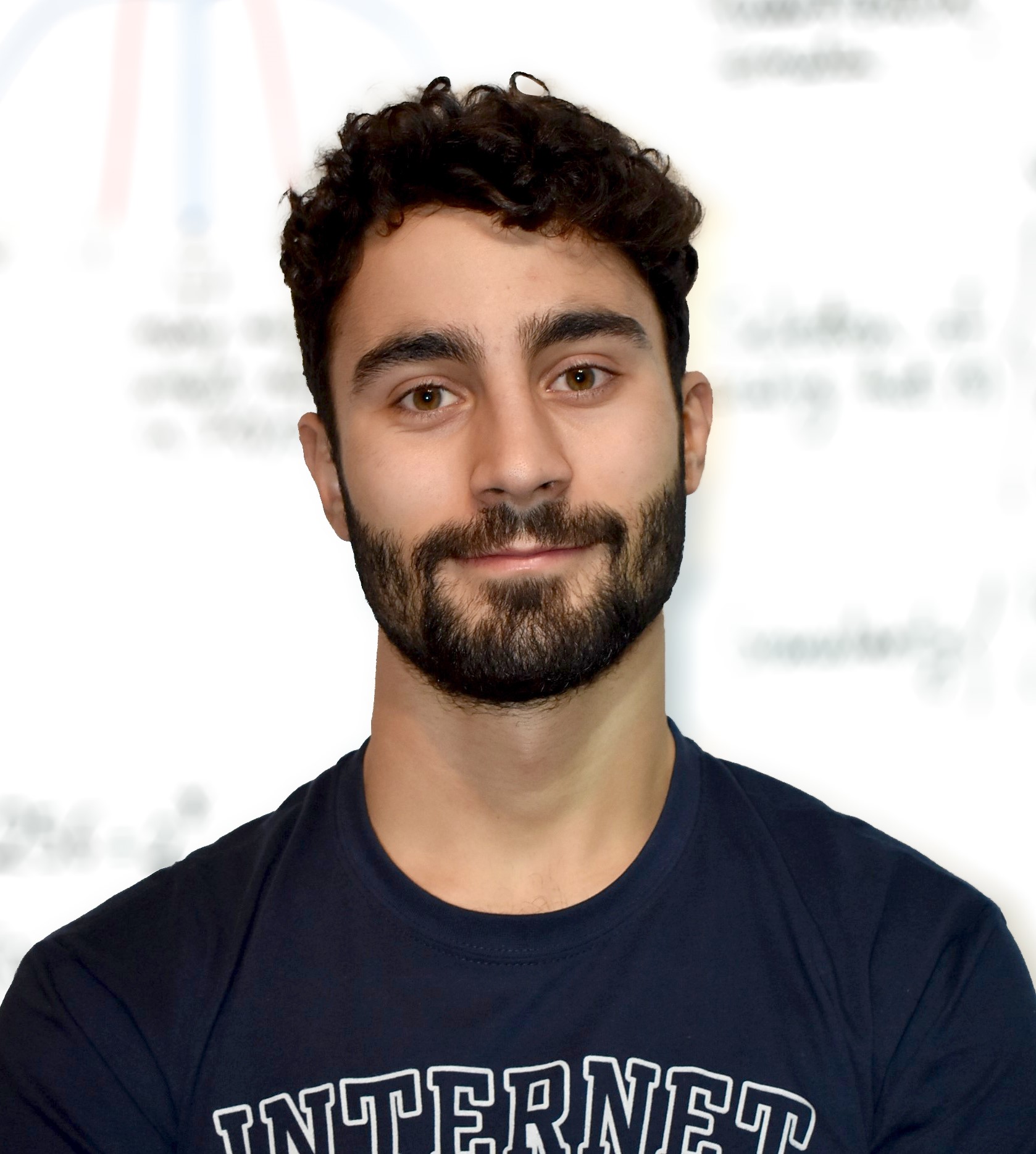}}]{Raul Murillo}
received a BSc in Computer Science and a BSc in Mathematics in 2019 from the Complutense University of Madrid (UCM), where he also earned an MSc in Computer Science in 2021 and a PhD in Computer Science in 2024. He was also a Visiting Researcher at Politecnico di Milano and the Barcelona Supercomputing Center. From 2022 to 2024 he was a Teaching Assistant of Computer Science at UCM, and he is currently working as hardware design engineer in the industry. His research interests include Computer Arithmetic, Computer Architecture, Approximate Computing, and Deep Neural Networks. 
\end{IEEEbiography}

\begin{IEEEbiography}[{\includegraphics[width=1in,height=1.25in,clip,keepaspectratio]{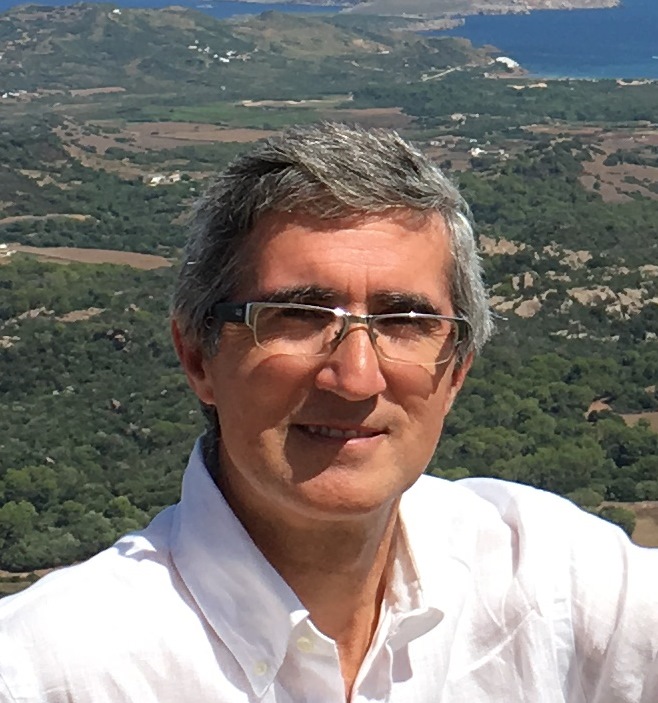}}]{Julio Villalba-Moreno} (Senior Member, IEEE)
received his B.S. degree in Physics from the University of Granada in 1986 and his Ph.D. in Computer Science from the University of Málaga, Spain. From mid-1986 to 1991, he worked as a design engineer in the R\&D Department of Fujitsu Spain. From 1986 to 2007, he was an Assistant Professor, and since 2007, he has been a Full Professor within the Department of Computer Architecture at the University of Málaga.

He was a Visiting Scholar in the Department of Electrical Engineering and Computer Science at the University of California, Irvine for one year. He served as an Associate Editor for IEEE Transactions on Computers from July 2011 to June 2015. He is member of the Steering Committee of the IEEE Symposium on Computer Arithmetic since 2015 and served as General Co-Chair of the 31st IEEE International Symposium on Computer Arithmetic (ARITH 2024).
\end{IEEEbiography}

\begin{IEEEbiography}[{\includegraphics[width=1in,height=1.25in,clip,keepaspectratio]{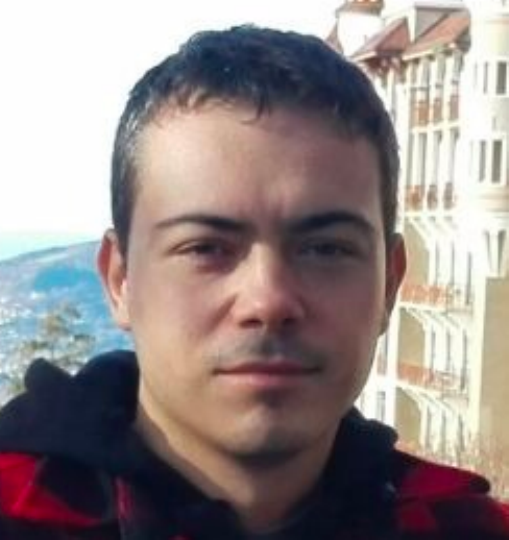}}]{Alberto A. Del Barrio} (Senior Member, IEEE)
received the Ph.D. degree in Computer Science from the Complutense University of Madrid (UCM), Madrid, Spain, in 2011. He has visited Northwestern University, University of California at Irvine and University of California at Los Angeles. Since 2021, he is an Associate Professor of Computer Science with the Department of Computer Architecture and System Engineering, UCM. His main research interests include Design Automation, Next Generation Arithmetic and Quantum Computing. Dr. del Barrio has been the PI of the PARNASO project, funded by the Leonardo Grants program by Fundación BBVA, and currently, he is the PI of the ASIMOV project, funded by the Spanish MICINN. These two projects have had a strong focus on posit arithmetic. Since 2019 he is an IEEE Senior Member and since December 2020, he is an ACM Senior Member, too.
\end{IEEEbiography}

\begin{IEEEbiography}[{\includegraphics[width=1in,height=1.25in,clip,keepaspectratio]{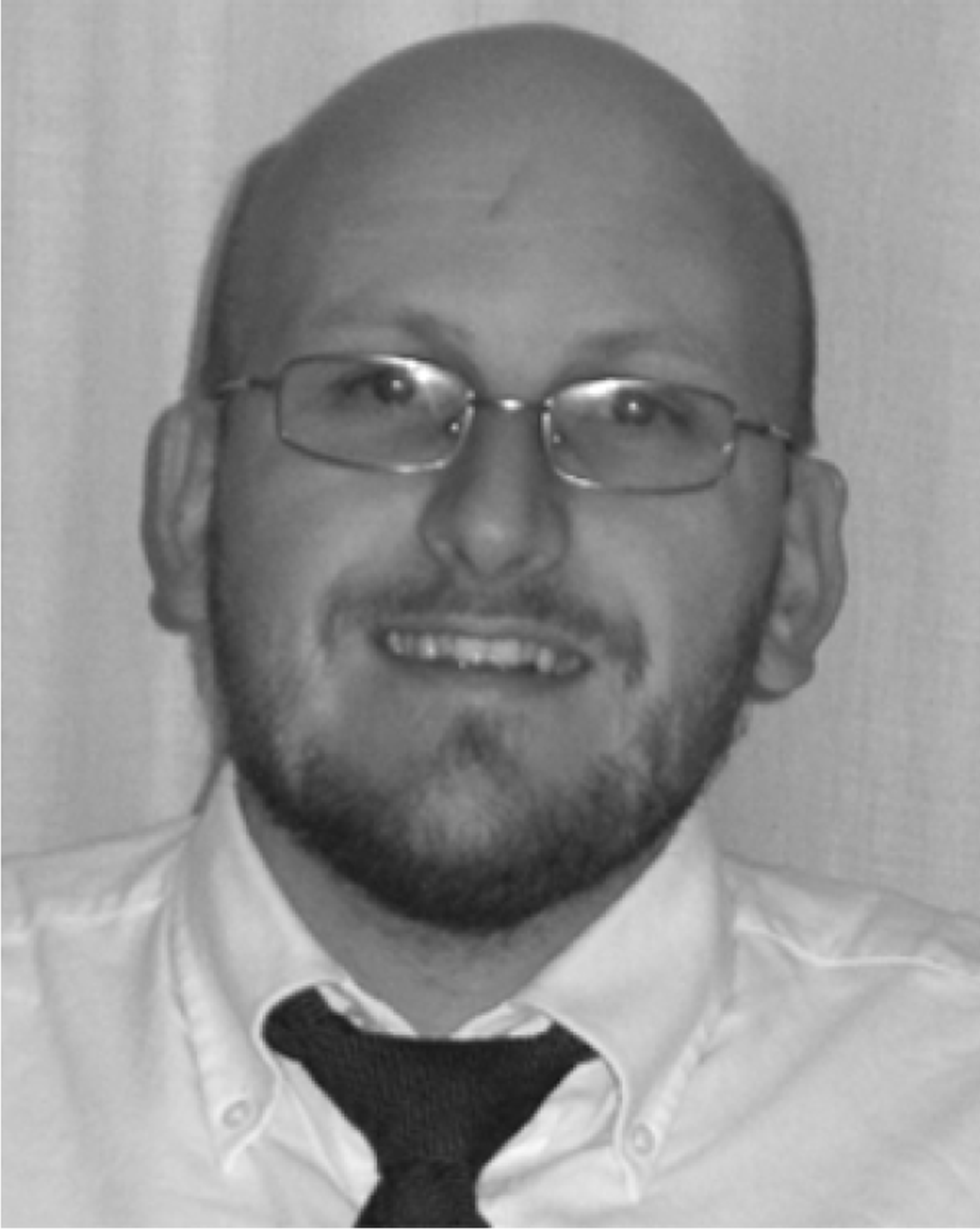}}]{Guillermo Botella} (Senior Member, IEEE)
received MSc degrees in Physics (1999) and Electronic Engineering (2002), and a Ph.D. in Physics from the University of Granada (2007), where he developed a hardware-accelerated implementation of a robust optical flow model. He was a research fellow funded by competitive fellowships, including EU Marie Curie and José Castillejo (Spanish government), working at the University of Granada, the Vision Research Laboratory at University College London, and FAMU-FSU College of Engineering (Florida). Since 2004, he has been with the Department of Computer Architecture and Automation at Complutense University of Madrid, becoming Full Professor in 2024. His research focuses on signal processing acceleration, computer arithmetic, and emerging paradigms such as neuromorphic, analog, and quantum computing. He has led contracts with IBM (quantum computing) and Banco Santander (HPC for microscopy). He is Deputy Editor-in-Chief of Digital Signal Processing (Elsevier). Awards include Best Paper IEEE CEDA Spain (2023) and Best PhD Thesis SARTECO (2024).

\end{IEEEbiography}



\vfill

\end{document}